# Borehole acoustic full-waveform inversion

Huaigu Tang [1,2*], Arthur Chuen Hon Cheng [1], Elita Yunyue Li [1], Xinding Fang [2]
[1] National University of Singapore
[2] Southern University of Science and Technology

## Summary

Full-waveform inversion (FWI) is a technique having the potential for building high-resolution elastic velocity models. We proposed to apply this technique to wireline monopole acoustic logging data to obtain the near wellbore formation velocity structures, which can be used in wellbore damage or fluid intrusion evaluation. A 2D FWI using monopole acoustic logging data is presented. The FWI is established in cylindrical coordinates instead of Cartesian coordinates in order to adapt to the borehole geometry. A preconditioner is designed for suppressing the influence of the strong borehole guided waves in the inversion. Synthetic tests demonstrate that high-resoultion elastic velocity profile around borehole can be inverted from monopole acoustic logging data by using the proposed method.

## Introduction

Formation compressional wave (P-wave) and shear wave (S-wave) velocities are important petrophysical properties as they are closely related to formation lithology (Bourbié et al., 1992.). Elastic wave velocity models are also important for seismic imaging (Yan and lines, 2001). Acoustic logging (Tang and Cheng, 2004) provides a reliable way to measure the downhole formation elastic wave velocities, which are useful information for lithology interpretation and also are the necessary inputs for well-to-siesmic ties.

Conventionally, acoustic logging is used to measure velocity variations along the borehole axial direction. Damage of wellbore and fluid intrusion can change the wellbore formation elasticity and thus affect the elastic wave velocities measured from acoustic logging (Winkler, 1997; Fang et al., 2014). Knowledge about the radial variations in near wellbore formation velocities can help recover the virgin formation velocities, and also provide important geomechanical information that is useful for well drilling and completion analysis. The information about radial velocity changes cannot be obtained from the conventional semblance analysis (Kimball and Mazetta, 1984), which is the standard method for extracting formation velocities in the oil industry. Hornby (1993) showed the potential of using ray-tracing tomography to obtain the radial velocity variations around borehole. Because FWI can achieve higher resolution than ray-based tomography; and FWI does not require travel time picking. The FWI method has potential to perform better than traditional traveltime tomography in borehole velocity inversion.

FWI is originally developed for seismic velocity model building. Because the propagation of both seismic wave and borehole acoustic wave is governed by the elastodynamic wave equation, the FWI method for acoustic logging data can adopt the workflow for seismic data. A 2D gradient descent FWI algorithm based on the borehole cylindrical coordinates is developed for monopole acoustic logging data. Effects of strong borehole guided waves, such as Stoneley wave and Pseudo Rayleigh waves (Cheng and Toksöz., 1981), are suppressed by weighting of gradients using a diagonal preconditioner. Starting models in the inversion are obtained from semblance analysis. Performannce and accuracy of the proposed inversion method is investigated through numerical synthetic tests.

## Method

The gradient descent inversion process is to minimize the following least-squares form objective function that represents the difference between the modeled data, which are the synthetic data calculated from forward modeling, and the measured acoustic logging data,

$$J = \frac{1}{2} \left\| {}^{mod}p - {}^{mes}p \right\|_2^2 \qquad (1)$$

where $J$ is the objective function, ${}^{mod}p$ is modeled data, ${}^{mes}p$ is measured data.

With the given measured data and starting velocity model, the true velocity model can be obtained by following the gradient descent inversion workflow shown in Figure 1. In each iteration, there are five key steps: forward modeling, data residue calculation, adjoint modeling, gradient calculation, and model updating.

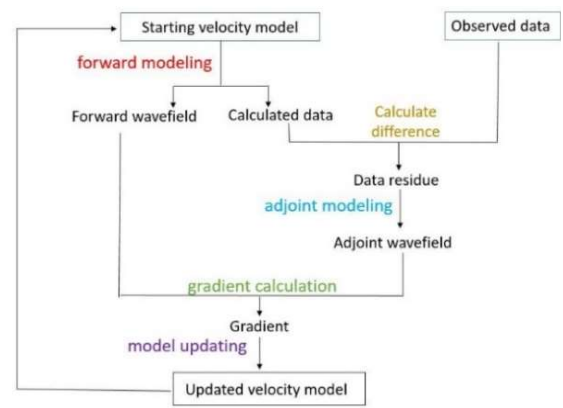

Figure 1. Workflow of gradient descent FWI method.

# Borehole acoustic FWI

Forward wavefields are obtained by forward borehole elastic wave propagation simulations. When the material properties are azimuthally invariant in a borehole model, the elastic wave equations in cylindrical coordinates for a monopole source can be expressed as (Wang and Tang, 2003),

$$\begin{aligned}
\frac{\partial v_r}{\partial t} &= \frac{1}{\rho}\frac{\partial \tau_{rr}}{\partial r} + \frac{1}{\rho}\frac{\partial \tau_{rz}}{\partial z} + \frac{1}{\rho}\left(\frac{\tau_{rr}}{r} - \frac{\tau_{\theta\theta}}{r}\right) \\
\frac{\partial v_z}{\partial t} &= \frac{1}{\rho}\frac{\partial \tau_{rz}}{\partial r} + \frac{1}{\rho}\frac{\partial \tau_{zz}}{\partial z} + \frac{1}{\rho}\frac{\tau_{rz}}{r} \\
\frac{\partial \tau_{rr}}{\partial t} &= (\lambda+2\mu)\frac{\partial v_r}{\partial r} + \lambda\frac{v_r}{r} + \lambda\frac{\partial v_z}{\partial z} + f \\
\frac{\partial \tau_{\theta\theta}}{\partial t} &= (\lambda+2\mu)\frac{v_r}{r} + \lambda\frac{\partial v_r}{\partial r} + \lambda\frac{\partial v_z}{\partial z} + f \\
\frac{\partial \tau_{zz}}{\partial t} &= \lambda\frac{\partial v_r}{\partial r} + \lambda\frac{v_r}{r} + (\lambda+2\mu)\frac{\partial v_z}{\partial z} + f \\
\frac{\partial \tau_{rz}}{\partial t} &= \mu\frac{\partial v_r}{\partial z} + \mu\frac{\partial v_z}{\partial r}
\end{aligned} \quad (2)$$

where $v_r$ and $v_z$ are the particle velocities in the r- and z-directions, respectively, $\tau_{rr}$, $\tau_{\theta\theta}$, and $\tau_{zz}$ are the normal stresses in the $r$-, $\theta$-, and $z$-directions, respectively, $\tau_{rz}$ is the shear stress in the $r$-$z$ plane, $f$ is the source function, $\rho$ is density, $\lambda$ and $\mu$ are formation elastic moduli. During the process of FWI, $\rho$ is treated as a known parameter as it can be obtained from density logs. $\lambda$ and $\mu$ are related to the P- and S-wave velocities as

$$\lambda = \rho V_P^2 - 2\rho V_S^2 \quad (3)$$
$$\mu = \rho V_S^2 \quad (4)$$

where $V_P$ and $V_S$ are respectively the P- and S-wave velocities. The elastic wave equation (1) is numerically solved by using a finite difference method (Cheng et al., 1995). Perfectly-match-layer (PML) absorbing boundary condition (Zhang and Shen, 2010) is applied on top, bottom and radial boundaries; axial symmetric boundary condition is applied on the borehole central axis.

Acoustic logs are the acoustic wave pressure field recorded in the borehole fluid. Therefore, $^{mod}p$ is obtained by recording the values of the normal stress (normal stresses are equal in a fluid) at the receiver positions in forward modeling. Further, Data residue, $\Delta p$, is obtained by subtracting the measured pressure data from the modeled data.

$$\Delta p = {}^{mod}p - {}^{mes}p \quad (5)$$

Then the adjoint wavefields are obtained by solving the following adjoint wave equations (Plessix, 2006) with the data residue as the source time function:

$$\begin{aligned}
\frac{\partial {}^Av_r}{\partial t} &= \frac{\partial\left[(\lambda+2\mu){}^A\tau_{rr}\right]}{\partial r} + \frac{\partial\left(\lambda{}^A\tau_{\theta\theta}\right)}{\partial r} + \frac{\partial\left(\lambda{}^A\tau_{zz}\right)}{\partial r} \\
&\quad + \frac{\partial\left(\mu{}^A\tau_{rz}\right)}{\partial z} - \frac{\lambda{}^A\tau_{rr}}{r} - \frac{(\lambda+2\mu){}^A\tau_{\theta\theta}}{r} - \frac{\lambda{}^A\tau_{zz}}{r} \\
\frac{\partial {}^Av_z}{\partial t} &= \frac{\partial\left(\lambda{}^A\tau_{rr}\right)}{\partial z} + \frac{\partial\left(\lambda{}^A\tau_{\theta\theta}\right)}{\partial z} + \frac{\partial\left[(\lambda+2\mu){}^A\tau_{zz}\right]}{\partial z} + \frac{\partial\left(\mu{}^A\tau_{rz}\right)}{\partial r} \\
\frac{\partial {}^A\tau_{rr}}{\partial t} &= \frac{\partial\left({}^Av_r/\rho\right)}{\partial r} - \frac{{}^Av_r/\rho}{r} + \Delta p \\
\frac{\partial {}^A\tau_{\theta\theta}}{\partial t} &= \frac{{}^Av_r/\rho}{r} + \Delta p \\
\frac{\partial {}^A\tau_{zz}}{\partial t} &= \frac{\partial\left({}^Av_z/\rho\right)}{\partial z} + \Delta p \\
\frac{\partial {}^A\tau_{rz}}{\partial t} &= \frac{\partial\left({}^Av_r/\rho\right)}{\partial z} + \frac{\partial\left({}^Av_z/\rho\right)}{\partial r} - \frac{{}^Av_z/\rho}{r}
\end{aligned} \quad (6)$$

where ${}^Av_r$, ${}^Av_z$, ${}^A\tau_{rr}$, ${}^A\tau_{\theta\theta}$, ${}^A\tau_{zz}$ and ${}^A\tau_{rz}$ are components of the adjoint wavefields.

By calculating the cross-correlation of the forward wavefields and the adjoint wavefields, we can obtain the gradients of model parameters $\lambda$ and $\mu$ as,

$$\begin{aligned}
\frac{\partial J}{\partial \lambda} &= \int_t \left[\left(\frac{\partial v_r}{\partial r} + \frac{v_r}{r} + \frac{\partial v_z}{\partial z}\right) \cdot \left({}^A\tau_{rr} + {}^A\tau_{\theta\theta} + {}^A\tau_{zz}\right)\right] dt \\
&= \int_t \left[(\nabla \cdot \mathbf{v}) \cdot \left({}^A\tau_{rr} + {}^A\tau_{\theta\theta} + {}^A\tau_{zz}\right)\right] dt
\end{aligned} \quad (7)$$

$$\frac{\partial J}{\partial \mu} = \int_t \left[2\frac{\partial v_r}{\partial r}{}^A\tau_{rr} + 2\frac{v_r}{r}{}^A\tau_{\theta\theta} + 2\frac{\partial v_z}{\partial z}{}^A\tau_{zz} + \left(\frac{\partial v_x}{\partial z} + \frac{\partial v_z}{\partial r}\right){}^A\tau_{rz}\right] dt \quad (8)$$

Then the model parameters can be updated by

$$\lambda = \lambda - \alpha\frac{\partial J}{\partial \lambda} \quad (9)$$
$$\mu = \mu - \alpha\frac{\partial J}{\partial \mu} \quad (10)$$

where $\alpha$ is a parameter controlling the updating step. $\alpha$ can be used for controlling the greatest change of the model parameters. The gradients should also be weighted by using preconditioner to suppress unwanted features in the gradients. The preconditioning scheme is explained in the next section.

Finally, the updated P- and S-wave velocity models can be calculated using the expressions below:

$$V_P = \sqrt{\frac{\lambda+2\mu}{\rho}} \quad (11)$$
$$V_S = \sqrt{\frac{\mu}{\rho}} \quad (12)$$

The velocity model is updated iteratively during the inversion until the data residue is smaller than a given threshold value.

# Borehole acoustic FWI

## Numerical Examples

Two numerical models are constructed in this study: (1) a layered-velocity model for illustrating the application of preconditioning and starting velocity model building based on semblance analysis; (2) a heterogeneous model with velocity variations in both the axial and radial directions for testing the applicability of the FWI algorithm.

The data acquisition geometry in the simulations is designed by following the setup of the Halliburton's modern acoustic logging tool. As shown in Figure 2, there are thirteen receiver stations with 0.5ft spacing. The tool has four monopole sources, but only the lower-near monopole, far monopole and ultra-far monopole are modelled in the simulations. So three measured datasets corresponding to the three monopole sources are generated at each logging depth and the tool logs every 0.5ft along the borehole axis. For convenience, recording geometry are further realigned to one monopole source and thirty nine receivers. Positions of the realigned source and receivers are shown in the central axis of the models in Figure 3.

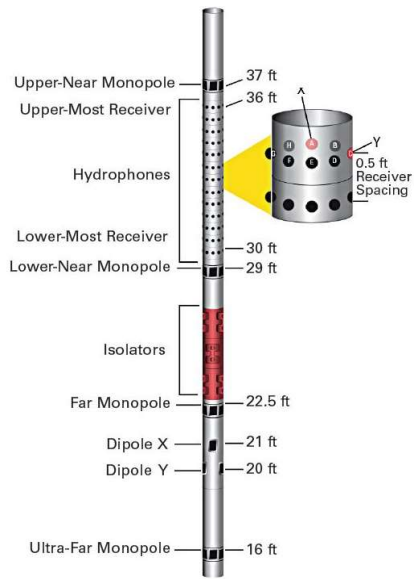

Figure 2. Schematic of the Halliburton Xaminer® Sonic Imager. (Walker et al., 2015)

Figure 3 shows the inversion results of the layered-velocity model. The starting velocity model is constructed based on semblance analysis (Figure 4). The velocities extracted from the semblance results of near-monopole data are taken as the formation velocities at the wellbore, while the velocities calculated from the semblance results of ultra-far monopole data are taken as the velocities for the formation at one meter away from wellbore. Then the velocities in between are obtained through linear interpolation. The finally obtained starting model is a smooth velocity model, as shown in the first column of Figure 3.

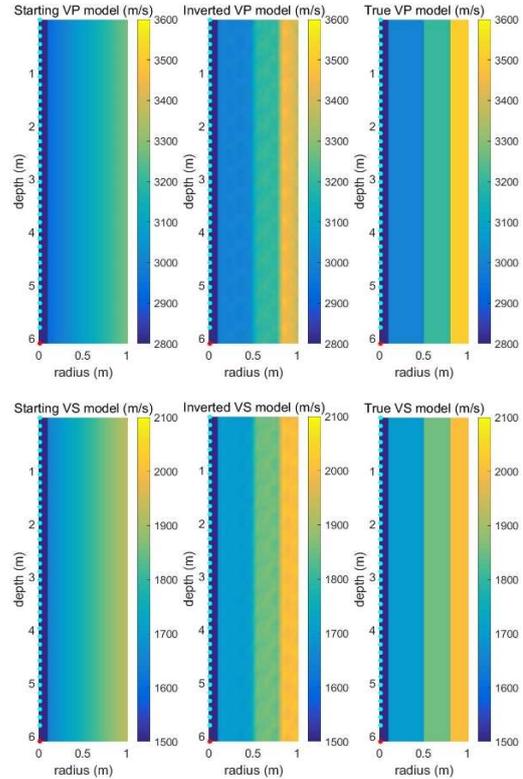

Figure 3. Borehole velocity models with three layers. The velocity profiles shown in the first, second and third columns respectively correspond to the starting model, inverted model and true model. Borehole radius is 10cm. Borehole fluid is water. Red dot indicates source position, blue dots indicate receiver positions.

In order to obtain the inverted model shown in the second column of Figure 3, it is necessary to pre-condition the gradient. This is because wellbore is a fluid-to-solid boundary, generating strong guided waves travelling along the wellbore (Figure 5). The borehole guided waves can result in abnormally large gradient near the wellbore. Therefore, a diagonal preconditioner is used for weighting the gradients and suppressing the effects of the guided waves (Figure 6).

Finally, a more realistic model containing low velocity flushed zone is used to validate the applicability of FWI (Figure 7). The model has velocty viarations in both radial and axial directions. The velocity profile of the inverted model is very close to that of the true model and the boundary between the flushed zone and the virgin formation can be clearly delineated in the inverted model.

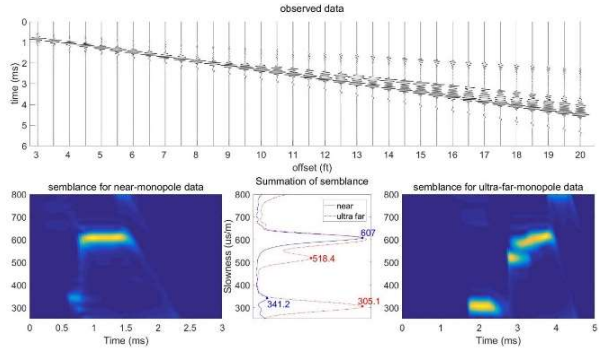

Figure 4. Semblance results of the measured data. Slowness (i.e. the inverse of velocity) of P- and S-wave obtained from the near-monopole data are respectively 341.2μs/m (2930 m/s) and 607.0 μs/m (1647 m/s). Slowness of P- and S-wave obtained from the ultra-far-monopole data are respectively 305.1 μs/m (3278 m/s) and 518.4 μs/m (1929m/s).

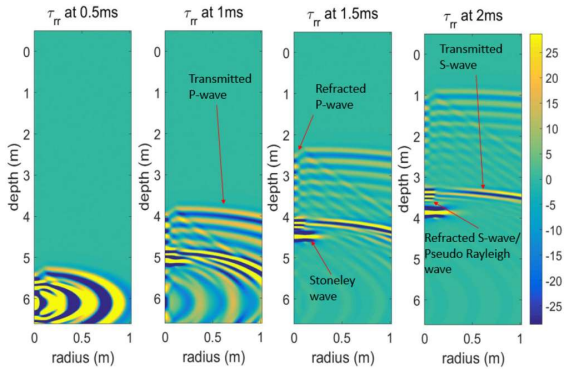

Figure 5. Radial normal stress wavefield in starting model.

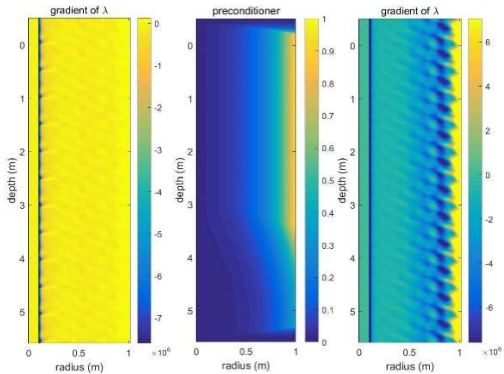

Figure 6. Gradient of $\lambda$ before preconditioning (left); the diagonal preconditioner (middle); and gradient of $\lambda$ after preconditioning (right).

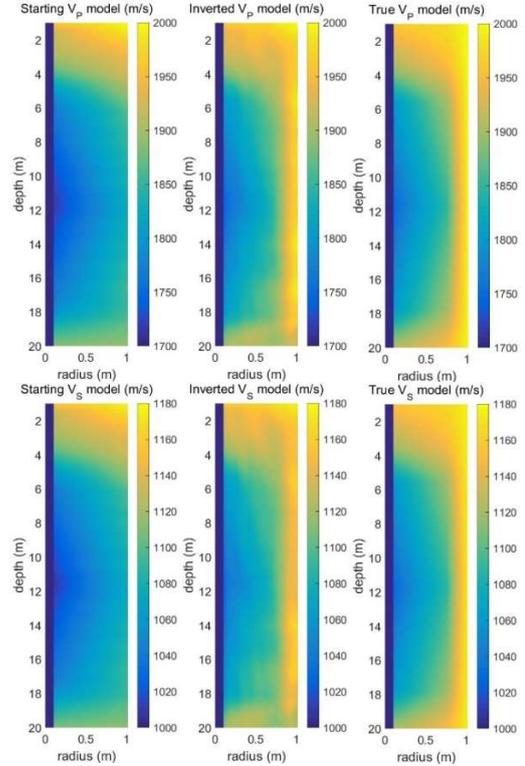

Figure 7. Borehole model with flushed zone. The velocity profiles shown in the first, second and third columns respectively correspond to the starting model, inverted model and true model. Radius of the borehole is 10cm.

**Conclusions**

A borehole acoustic FWI method using monopole acoustic logging data has been developed for near wellbore velocity inversion. This method is established by modifying the conventional seismic FWI workflow from Cartesian coordinates to borehole cylindrical coordinates. Starting models are constructed based on semblance analysis. Diagonal preconditioning is applied to suppress the influence of borehole guided waves in gradient calculation. Our numerical tests have demonstrated that the proposed FWI method is applicable to borehole models that exhibit velocity variations in both the axial and radial directions.

**Acknowledgements**

The authors acknowledge the EDB Petroleum Engineering Professorship for financial support. This work is also supported by the National Natural Science Foundation of China (grant no. 41704112)

# Borehole acoustic FWI